\documentclass[journal=mamobx,manuscript=article,layout=twocolumn]{achemso}

\usepackage{amsthm}
\usepackage{mathtools}
\usepackage{physics}
\usepackage{xcolor}
\usepackage{graphicx}
\usepackage{csquotes}
\usepackage{amsmath}
\usepackage{amssymb}
\usepackage{bm}
\usepackage{caption}
\usepackage{subcaption}
\usepackage[english]{babel}
\usepackage{algorithm}
\usepackage{algpseudocode}
\usepackage{cuted}
\usepackage{chemfig}
\usepackage{array}
\usepackage{sectsty}

\allsectionsfont{\raggedright}

\usepackage{placeins}
\let\Oldsection\section
\renewcommand{\section}{\FloatBarrier\Oldsection}
\let\Oldsubsection\subsection
\renewcommand{\subsection}{\FloatBarrier\Oldsubsection}
\let\Oldsubsubsection\subsubsection
\renewcommand{\subsubsection}{\FloatBarrier\Oldsubsubsection}

\usepackage[version=3]{mhchem} 

\usepackage[colorlinks,allcolors=cyan!70!black]{hyperref} 

\title{Procedural Construction of Atomistic Polyurethane Block Copolymer Models for High Throughput Simulations}

\author{Dominic Robe}
\affiliation{Department of Mechanical Engineering, Faculty of Engineering and Information Technology, The University of Melbourne, 700 Swanston St., Melbourne, Victoria, Australia}

\author{Adrian Menzel}
\affiliation{Platforms Division, Defence Science and Technology Group, Port Melbourne, Victoria, Australia}
\author{Andrew W Phillips}
\affiliation{Platforms Division, Defence Science and Technology Group, Port Melbourne, Victoria, Australia}
\author{Elnaz Hajizadeh}\email{ellie.hajizadeh@unimelb.edu.au}
\affiliation{Department of Mechanical Engineering, Faculty of Engineering and Information Technology, The University of Melbourne, 700 Swanston St., Melbourne, Victoria, Australia}

\date{January 2024}

\begin{document}

\abstract{In this work, methods are presented to automatically generate a fully atomistic LAMMPS models of arbitrary linear multi-block polyurethane copolymers. The routine detailed here receives as parameters the number of repeat units per hard block, the number of units in a soft block, and the number of soft blocks per chain, as well as chemical formulae of three monomers which will form the hard component, soft component, and chain extender. A routine is detailed for converting the chemical structure of a free monomer to the urethane bonded repeat units in a polymer. The python package RadonPy is leveraged to assemble these units into blocks, and the blocks into copolymers. Care is taken in this work to ensure that plausible atomic charges are assigned to repeat units in different parts of the chain. The static structure factor is calculated for a variety of chemistries, and the results compared with wide angle x-ray scattering data from experiments with corresponding composition. The generated models reproduce the amorphous halo observed in the scattering data as well as some of the finer details. Structure factor calculations are decomposed into the partial structure factors to interrogate the structural properties of the two block types separately. Parametric surveys are carried out of the effects of various parameters, including temperature, soft block length, and block connectivity on the observed structure. The routine detailed here for constructing models is robust enough to be executed automatically in a high throughput workflow for material design and discovery.}

\maketitle

\section{Introduction}

Modern polymer synthesis resources make it possible to compose copolymer materials with diverse repeat unit types and sequences\cite{Akindoyo2016-ms,Cheng2022-qm,Petrovic1991-vi}. While effective copolymer products have been established for many applications, it is often not certain what composition would optimize material performance for a new application, or if better formulations exist somewhere in the vast design space. Tools like machine learning and intelligent sampling can help to explore this space\cite{Pugar2020-ba,Ethier2023-qd,Weeratunge2022-dm,Pugar2022-cg, aplc, el}, but carrying out experiments is still a rate-limiting step in the exploration process. Simulations have been limited as a tool for surveying different chemistries due to the need for an expert to construct an atomistic model of a particular material. Construction of these models involves meticulous specification of atom types and dozens of force field parameters. In the same way that a robotic synthesizer enables high-throughput experiments, a robust platform for procedurally generating atomistic models would enable rapid exploration of the polymer design space.

The recently developed Python package RadonPy\cite{hayashi2022radonpy} has made it possible to automatically deduce atom types and identify force field parameters for atomistic models of many compounds. This package provides tools to specify a set of monomer chemistries, then hierarchically construct polymer models with arbitrary sequences of those monomers. RadonPy leverages the quantum field theory package Psi4\cite{10.1063/5.0006002} to assign charges to each atom in a monomer. This tool has unlocked a new domain of high-throughput atomistic polymer simulations.

While RadonPy is a powerful tool, there are some notable challenges remaining before this tool can be applied to a particular design task. One is that the chemical structure of a repeat unit might be slightly different at chain ends or a bond between blocks, as opposed to units within a block. A procedure for constructing copolymer models must account for these heterogeneities. In this work we implement a routine to handle these linkages automatically for arbitrary polyurethane chemistries. The importance of polyurethanes as a demonstration of this sort of routine will be discussed below.

Second, the workflow suggested by the RadonPy documentation calculates the charge distribution on an isolated monomer. While the full context of the polymer chain would likely yield only a small correction, the presence of the adjacent units in the chain will certainly influence the charges on the atoms at the edges of the unit. Particularly in the case of a chemistry like poly(tetramethylene oxide) (PTMO), assigning charges to a solitary monomer can yield a charge distribution with incorrect symmetry. The backbone of a PTMO chain may be represented as "...OCCCCOCCCCOCCCCO..." Here the central four carbon atoms should clearly have a symmetrical charge distribution, but if the repeat unit is represented as "OCCCC", then the charge distribution on the carbons will be asymmetrical. This particular example could be mitigated by representing the repeat unit as "CCOCC", but the charges on first and last carbon atoms will still be incorrect due to the missing context of their neighbouring carbon atoms in the chain. This problem is complicated by the variety of possible neighbourhoods for a monomer. It could be within a block, at a block transition, or at a chain end. There are also the edge cases of a blocks or chains composed of a single monomer. In this work we implement a routine to enumerate the possible neighborhoods for each type of monomer and calculate the charge distribution for each situation.

Finally, generated models must be validated against experimental measurements to ensure that the series of routines used to construct the models are capable of producing trustworthy results. In this work, we measure the static structure factor in simulation, and compare the results to wide angle x-ray scattering (WAXS) data\cite{SUN2006650,Buckley2010-ni} for a variety of polyurethane compositions and chemistries. We further present breakdowns of the partial structure factors, which provide insight about the microstructure that is not available to experiment. We then demonstrate the flexibility of the model generator by performing surveys of the effect of various composition parameters on the structure factor.

The broader motivation for this implementation and validation effort should also be mentioned. Simulation models of polymers exist at many scales of resolution\cite{Park2021-vq,Lempesis2017-kh,Speckhard1987-kj,Gallu2020-iq,Hu2016-lt,Dunbar2020-nm,Xia2019-ai,Xia2018-fy, MD1, MD2, MD3, MD4, POP, MD5}. These models involve material parameters that often cannot be derived directly from real synthesis variables or measured properties. These model parameters tend to be used as fitting parameters to align model results with experimental observation, then infer the microscopic character of the real system from the model. This approach is informative if a model effectively encodes the relevant physics in meaningful parameters\cite{Giuntoli2021-hu}. However, it is always possible for such models to disregard an unknown phenomenon, yielding a poor fit, or to yield a result that looks appropriate macroscopically, but for a different reason than in the real system. These situations can be difficult to diagnose in the absence of microscopic information. The purpose of atomistic simulations then is to provide bottom-up information grounded in a realistic representation of a material\cite{Shireen2022-hm}. That information could be employed in a number of ways\cite{Ahn2023-qt}, including as training data for machine learning and optimization algorithms\cite{Weeratunge2022-dm,hayashi2022radonpy,dabiri2023fractional}, or as a fitting target to constrain coarse grained model parameters\cite{Weeratunge2023,Song2018-xt}. These use cases are presently limited in throughput by the time consuming step of an expert manually constructing a bespoke atomistic model for any chemistry of interest. Many atomistic molecular dynamics simulations have been reported on, which are limited to a particular chemistry\cite{Lempesis2017-kh,Hsieh2021-jw,Zhu2018-jt,Zhu2018-rr,Repakova2004-rc,Rahmati2012-dd,Shireen2023-rj,Mirhosseini2022-wf,Lu2021-sh,Park2021-vq}. In this work, we demonstrate that useful atomistic models can be constructed automatically with even the choice of component chemistries as parameters.

\section{Polyurethane Block Layout}

\begin{figure*}
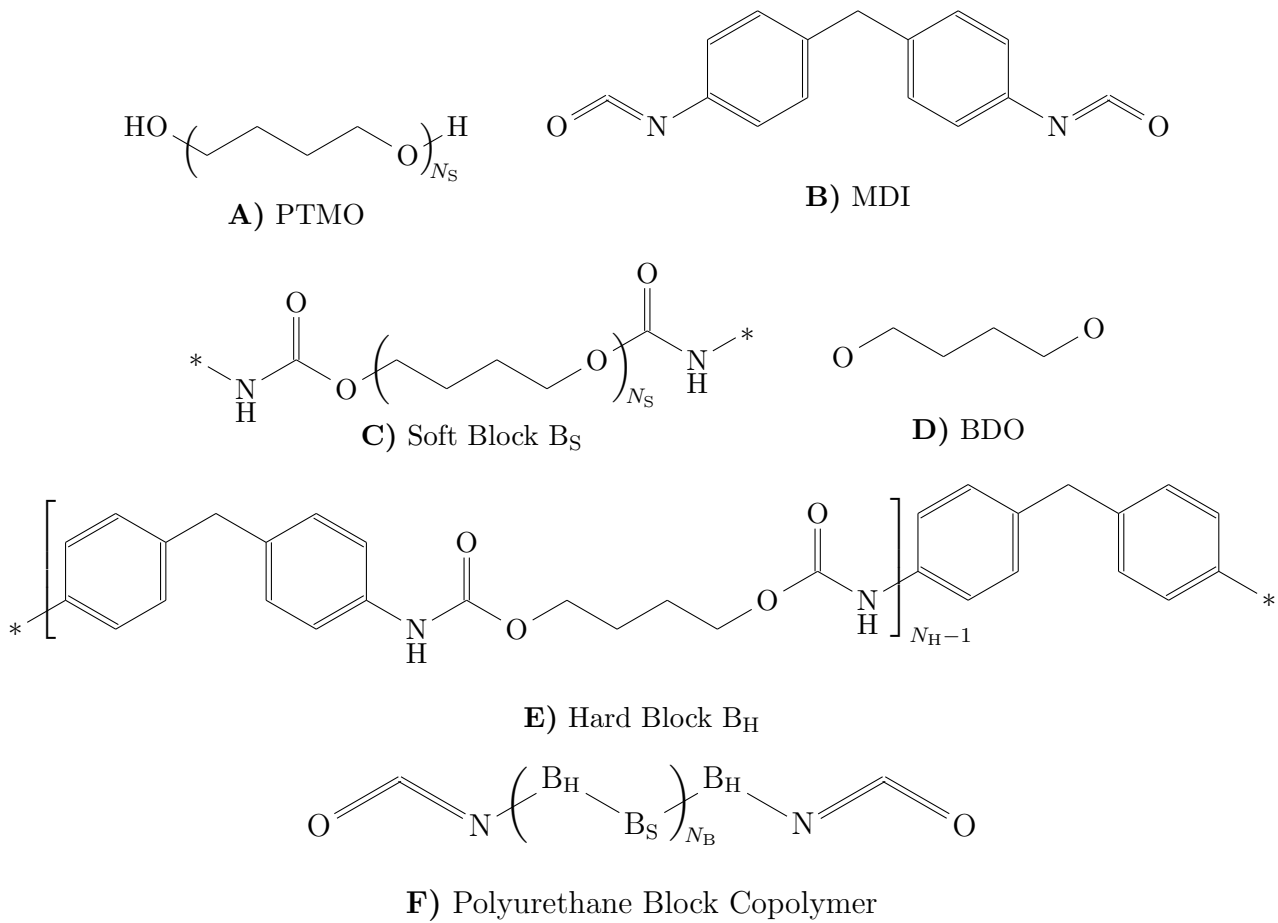


\small
\setchemfig{atom sep=2em}
\chemname[3ex]{\chemfig[angle increment=30]{HO-[@{PTMOhead1,.5}-1]-[1]-[-1]-[1]-[-1]O-[@{PTMOtail1,.3}1]H}}{\textbf{A)} PTMO}
\polymerdelim[height=5pt,indice=\!\!N_\mathrm{S}]{PTMOhead1}{PTMOtail1}
\qquad
\chemname{\chemfig[angle increment=30]{O=[1]=[-1]N-[1](-[3]=_[1]?[a])=^[-1]-[1]=^[3]?[a]-[1]-[-1](-[1]=_[-1]?[b])=^[-3]-[-1]=^[1]?[b]-[-1]N=[1]=[-1]O}}{\textbf{B)} MDI}
\vskip 1em
\chemname[2ex]{\chemfig[angle increment=30]{\ast-[-1]\chembelow{N}{H}-[1](=[3]O)-[-1]O-[@{PTMOhead2,.5}1]-[-1]-[1]-[-1]-[1]O-[@{PTMOtail2,.3}1](=[3]O)-[-1]\chembelow{N}{H}-[1]\ast}}{\textbf{C)} Soft Block $\mathrm{B_S}$}
\polymerdelim[height=10pt,indice=\!\!N_\mathrm{S}]{PTMOhead2}{PTMOtail2}
\qquad
\chemname{\chemfig[angle increment=30]{O-[1]-[-1]-[1]-[-1]-[1]O}}{\textbf{D)} BDO}
\vskip 1em
\chemname[3ex]{\chemfig[angle increment=30]{\ast-[@{HardBlockHead,.5}1](-[3]=_[1]?[a])=^[-1]-[1]=^[3]?[a]-[1]-[-1](-[1]=_[-1]?[b])=^[-3]-[-1]=^[1]?[b]-[-1]\chembelow{N}{H}-[1](=[3]O)-[-1]O-[1]-[-1]-[1]-[-1]-[1]O-[1](=[3]O)-[-1]\chembelow{N}{H}-[@{HardBlockTail,.5}1](-[3]=_[1]?[c])=^[-1]-[1]=^[3]?[c]-[1]-[-1](-[1]=_[-1]?[d])=^[-3]-[-1]=^[1]?[d]-[-1]\ast}}{\textbf{E)} Hard Block $\mathrm{B_H}$}
\polymerdelim[delimiters={[]},height=40pt,indice=N_\mathrm{H}-1]{HardBlockHead}{HardBlockTail}
\vskip 1em
\normalsize
\setchemfig{atom sep=3em}
\chemname{\chemfig[angle increment=30]{O=[1]=[-1]N-[@{MultiBlockHead,.5}1]B_{H}-[-1]B_{S}-[@{MultiBlockTail,.5}1]B_{H}-[-1]N=[1]=[-1]O}}{\textbf{F)} Polyurethane Block Copolymer}
\polymerdelim[indice=\!\!N_\mathrm{B}]{MultiBlockHead}{MultiBlockTail}
\caption{\label{fig:chemical formulas}The molecular structure of a polyurethane block copolymer in the present work. For reference, the unreacted forms of A) A PTMO prepolymer and B) unreacted MDI. C) Each soft block is composed of a soft prepolymer of the specified length, terminated by urethane bonds leading to hard blocks. D) BDO as an example chain extender E) Each hard block is composed of the specified number of hard monomers, alternating with the chain extender, linked by urethane bonds. F) The full structure of a constructed multiblock copolymer chain.}
\end{figure*}

To construct a model generator that is viable for automatic evaluation, it must handle consistently all of the possible design parameters. It is therefore vital at the start to articulate precisely the architecture of polyurethane (PU) block copolymers (BCPs) and their design variables. PUs consist of three types of monomer, which we will call the hard segment, soft segment, and chain extender. The soft segments are introduced as a "prepolymer" diol with arbitrary molecular weight as represented in Fig~\ref{fig:chemical formulas}A. The prepolymers are typically in the range 1000-2000~Da\cite{Petrovic1991-vi}, which corresponds to $N_\mathrm{S}=$15-30 monomers for PTMO. The hard component is a diisocyanate, such as 4,4'-methylene diphenyl diisocyanate (MDI) shown in Fig.~\ref{fig:chemical formulas}B. The hard component and soft prepolymers undergo a urethane reaction, yielding an "isocyanated" prepolymer. Fig.~\ref{fig:chemical formulas}C would represent the result if each asterisk represents the remainder of a hard segment, missing one isocyanate group which has been replaced by the urethane bond. Finally, the chain extender is introduced to join the isocyanated prepolymers into a multi-block copolymer. The chain extender is a short diol, such as 1,4-butanediol (BDO) in Fig.~\ref{fig:chemical formulas}D. The chain extender undergoes urethane reactions with the end groups of the isocyanated prepolymers. If there are more hard segment and chain extender molecules than prepolymers, then chain extenders and hard segments can form an alternating copolymer referred to as a "hard block", shown in Fig.~\ref{fig:chemical formulas}E. The hard component, chain extender, and soft prepolymers are combined in the stoichiometric ratio $N_\mathrm{H}:N_\mathrm{H}-1:1$ where $N_\mathrm{H}$ is the resulting number of hard monomers per hard block. If $N_\mathrm{H}$=1, no chain extender is used, and each hard monomer bonds directly to a prepolymer on either side. Fig.~\ref{fig:chemical formulas}F then represents the full BCP composed of $N_\mathrm{B}$ linked prepolymers which are now "soft blocks" and $N_\mathrm{B}+1$ hard blocks.

Fig.~\ref{fig:chemical formulas} implies that each soft block has exactly $N_\mathrm{S}$ monomers, each hard block has exactly $N_\mathrm{H}$ hard monomers, and each chain exactly $N_\mathrm{B}$ soft blocks, which is not true in a real material. The lengths of each block and the number of blocks in each chain will form a distribution determined by the kinetics of the methods used to synthesize the PU\cite{Speckhard1987-kj}. For the present work our model generation routine will assume these details are homogeneous, and take the design variables $N_\mathrm{H}$, $N_\mathrm{S}$, and $N_\mathrm{B}$ as parameters. The procedure could be extended to include variance parameters for these three quantities. Sophisticated techniques exist to model polymer reaction kinetics. Such methods could be used to generate detailed distributions for block and chain lengths.

Polyurethane block copolymers are a highly informative class of materials for which to validate the procedural model construction developed in this work. The modest complication of accounting for the urethane bonds demonstrates a format for specifying chemical components that allows a user to naturally provide lists of ingredients (such as diisocyanates). PU BCPs provide an interesting non-trivial yet comprehensible design space, the parameters of which are $N_\mathrm{H}$, $N_\mathrm{S}$, and $N_\mathrm{B}$, hard chemistry, soft chemistry, and extender chemistry. This leads to a class of materials with diverse properties and applications that are nevertheless possible to represent with a single generalized model. PUs are also a challenging material to model accurately because the specific geometries of the three components affect their compatibility\cite{Velankar2000-dd,Velankar2000-wk,Wang2022-qm,Petrovic1991-vi}. In particular, the urethane groups scattered along the hard blocks lead to hydrogen bonding between the hydrogen in one urethane group interacting with the oxygen in a urethane group in a neighbouring chain. The diversity of properties in PUs is often attributed to variation in the hydrogen bonding capacity of different combinations of chemistries\cite{Pugar2020-ba,Petrovic1991-vi,Qi2005-nv}. Some hard components and chain extenders are understood to frustrate these bonds, while others form crystal structures reinforced by the bonds. If there are additional donor or acceptor groups in any of the three components, this can also impact material properties.

Polyurethanes are also notable for their phase separated microstructures\cite{Velankar2000-dd,Velankar1998-bp,Cheng2022-qm,Gallu2020-iq,Dumais1985-ay,Leung1985-ok}. Because the hard and soft blocks are immiscible, they prefer to phase separate. But because the blocks are bonded together, the phase separated domains can only span a length scale around the block length. Depending on the lengths and compatibility of the blocks, the phase separated domains can take on a variety of morphologies\cite{Mishra2011-ik,Dhollander2010-dc,Hu2016-lt,Koberstein1992-ck,Gallu2020-iq,Ren2016-up,Korley2006-ds,James_Korley2006-lw}, which in turn affect the bulk material properties. Often the hard component occupies a lower volume fraction than the soft blocks, so the typical illustration is of crystalline hard domains embedded in a soft matrix\cite{SUN2006650,Pugar2020-ba}. However, these illustrations are inferred from experimental measurements that cannot directly observe the molecular arrangements. If valid representations of PU BCPs can be constructed procedurally, then molecular dynamics simulations will allow these structures to be observed directly.

One measurement that is commonly used to characterize PU BCP microstructures is wide angle x-ray scattering (WAXS)\cite{Wang2022-qm,SUN2006650,Dhollander2010-dc,witzleben2015investigation,Koberstein1992-gd,James_Korley2006-lw,Waletzko2009-ka}. This provides information about periodicities in molecular structures at the scale of 2-10~\AA. For instance, crystal structures are often identified by peaks in the WAXS profile at particular wave vectors, corresponding to lattice spacings. In disordered materials, an "amorphous halo" is observed as a broad, shallow peak representing the distribution of disordered intermolecular spacings. The WAXS scattering intensity can be compared to the static structure factor, which can be calculated from the atomic positions in a molecular dynamics simulation. Small angle x-ray and neutron scattering is also used to characterize phase separated morphologies on length scales in excess of 100~\AA\cite{Wang2022-qm,SUN2006650,Dhollander2010-dc,witzleben2015investigation,Phillips1988-uj,Waletzko2009-ka,Leung1985-ok,Miller1984-pl}, but these scales are challenging to capture routinely with atomistic MD simulations, so this work will focus on wide angle data. In addition to the total static structure factor, simulation data can be partitioned by block type to calculate the partial structure factors, which characterize structural scales among particular species. This disassembly of the structure factors provides additional clarity about which components are contributing to particular features of the scattering profile, which is unavailable in experiment.

\section{Routine for generating polyurethane chemical structures}

RadonPy\cite{hayashi2022radonpy} employs the data structures in the package RDkit to manage molecular structures and initial configurations. RDkit includes methods to initialize a molecule from a SMILES string. RadonPy builds on this functionality by allowing asterisks to be included at the beginning and end of a smiles string to represent polymerization points. So the string "*OCCCC*" is used to initialize a PTMO repeat unit. The locations of hydrogen atoms are inferred from the smiles string and are included explicitly in the generated unit.

RadonPy includes an assortment of methods for concatenating these units. Care must be taken by a user to ensure that the unit concatenation instructions correspond to the intended polymer. For instance, the soft block is terminated by oxygens at both ends, one of which is not accounted for by the repeat unit. This can be included by creating another unit with an appended oxygen, such as "*OCCCO*" for PTMO, which is used as the last soft monomer in a block.  

For the hard blocks, one must account for most of the hard units ending in urethane bonds, except for at the chain ends, where one isocyanate group remains. The smiles string for a diisocyanate can always take the form "O=C=N...N=C=O" where "..." represents the bulk of the hard unit. A urethane bonded hard segment can therefore be represented as "*C(=O)N...NC(=O)*" in RadonPy. Further, the terminal units can be encoded by only replacing one isocyanate group with the urethane bond in this manner, as RadonPy allows units with only one asterisk to terminate a chain. Chain extender units are always sandwiched between hard units, so they are straightforwardly represented by including asterisks before and after their normal SMILES string.

With these units declared, one can now construct a PU BCP. First, an alternating copolymer of $N_\mathrm{H}$ hard units and $N_\mathrm{H}$-1 chain extenders is generated. A variant is generated for the terminal hard blocks with one normal hard unit replaced with the terminal hard unit. Then a chain of $N_\mathrm{S}$-1 soft units and the terminal soft unit is generated. Finally, the full BCP is constructed by joining one terminal hard block, $N_\mathrm{B}$-1 pairs of soft and hard blocks, the last soft block, and the final terminal hard block.

\section{Routine for assigning charges}

RadonPy includes wrapper methods for the quantum field theory package Psi4\cite{10.1063/5.0006002}. A molecule's conformation is first established using force fields with all atoms being charge neutral. This conformation is then used to initialize a Psi4 computation to determine the charge distribution. In order to ensure that the charge distributions are appropriate for the neighborhood of a monomer, in this work, we consider all possible neighbourhoods for each unit type, and run the charge assignment method on each neighbourhood, and extract the charges for the central unit. For example, a soft unit at the edge of a soft block would have a different charge distribution than one in the middle of the block. If we note a hard unit as H, a soft unit as S, a chain extender as E, and a free chain end as x, then the possible neighbourhoods are enumerated in Table~\ref{tab:neighbours}. Note that a naive consideration of all possible permutations of three components would involve more than three times as many quantum simulations.

\begin{table*}
    \centering
    \begin{tabular}{p{16mm} |p{17mm} | p{17mm} | p{17mm} | p{17mm} }
                       & Middle Hard               & Middle Extender& Middle Soft                           &               \\
        \hline
        Preceding Unit &                           &                &                                        & Following Unit\\
        \hline
        \centering x   & \centering \underline{\underline{xHx}}&    & \centering \underline{\underline{xSx}} & \centering\arraybackslash x  \\
        \centering x   &                           &                &                                        & \centering\arraybackslash H  \\
        \centering x   & \centering \underline{xHE}&                &                                        & \centering\arraybackslash E  \\
        \centering x   & \centering \underline{xHS}&                & \centering \underline{\underline{xSS}} & \centering\arraybackslash S  \\
        \hline
        \centering H   &                           &                &                                        & \centering\arraybackslash x  \\
        \centering H   &                           & \centering HEH & \centering HSH                         & \centering\arraybackslash H  \\
        \centering H   &                           &                &                                        & \centering\arraybackslash E  \\
        \centering H   &                           &                & \centering \underline{HSS}             & \centering\arraybackslash S  \\
        \hline
        \centering E   &                           &                &                                        & \centering\arraybackslash x  \\
        \centering E   &                           &                &                                        & \centering\arraybackslash H  \\
        \centering E   & \centering EHE            &                &                                        & \centering\arraybackslash E  \\
        \centering E   & \centering EHS            &                &                                        & \centering\arraybackslash S  \\
        \hline
        \centering S   &                           &                & \centering \underline{\underline{SSx}} & \centering\arraybackslash x  \\
        \centering S   &                           &                & \centering \underline{SSH}             & \centering\arraybackslash H  \\
        \centering S   &                           &                &                                        & \centering\arraybackslash E  \\
        \centering S   & \centering SHS            &                & \centering SSS                         & \centering\arraybackslash S 
    \end{tabular}
    \caption{Neighbourhoods that must be evaluated for unique charge assignments. "x" indicates that the "middle" unit is at a chain end. Unfilled locations in the table indicate tripplets that do not occur in PU BCPs as detailed in this work. Underlined units involve a variation in the chemical structure. Specifically, for hard units at chain ends, urethane groups are replaced with isocyanate groups. For soft units at the end of a block, an additional oxygen is included. Double underlines indicate neighbourhoods that are only present in blends of hard and soft components rather than BCPs. }
    \label{tab:neighbours}
\end{table*}

As an illustration, two PU architectures are sufficient to demonstrate all of these neighborhoods. For $N_\mathrm{B}$=1, $N_\mathrm{S}$=3, and $N_\mathrm{H}$=3, we can represent a chain with the monomer sequence "xHEHEHSSSHEHEHx". Meanwhile, the chain with $N_\mathrm{B}$=2, $N_\mathrm{S}$=1, and $N_\mathrm{H}$=1 is represented as "xHSHSHx". This is not to suggest that $N_\mathrm{S}$=1 is a typical architecture, but that the model generation routine is valid in the edge cases of parameter space, which is vital for a platform to be interoperable with other software.

Additionally, it would be useful to be able to model hard "blocks" and soft "blocks" as separate chains in a blend. To allow this, a few more neighborhoods must be included for the cases of a solitary raw diisocyanate and the free ends of a soft prepolymer. These neighbourhoods are double underlined in Table~\ref{tab:neighbours}. Again, this is not to suggest that a practical material could involve an alternating copolymer of hard and chain extender units blended with soft polymers, but that our platform is capable of modelling such a material in a physically consistent way to investigate the effect of block connectivity systematically.

Each of these trimers (or in some cases dimers or lone monomers) can be generated and assigned charges, then the charges for the central unit extracted and stored. Some small net charge will have accumulated on the middle unit, so the positive and negative charges are scaled by the necessary small factor to ensure each unit is charge neutral. The files containing the charge distributions for each neighbourhood can be labelled with their constituent hard, soft, and extender chemistries, so that future simulations of the same components with different architecture or conditions don't need to repeat the non-trivial quantum field calculations.

\section{Results}

Sun, et al. \cite{SUN2006650} reported WAXS measurements of a polyurethane composed of PTMO soft segments, with MDI and BDO hard segments. The number average molecular weight of their PTMO prepolymers $M_\mathrm{n}$ was 2000 g/mol with a polymerization index of 2.25. Each hard segment monomer consisted of two MDI and one BDO chain extender. While the algorithm presented here for generating model polymer molecules does not allow a distribution of block lengths, the average characteristics of the material studied by Sun, et al. are straightforward to reproduce. As a validation of our modelling, we have simulated a system of PTMO, MDI, and BDO, with the $N_H=2$ and $N_S=31$. While this particular experimental report didn't investigate the total molecular weight of the synthesized block copolymers, PU BCPs can contain tens or hundreds of blocks with a wide distribution. We simply used $N_B=3$ to keep the number of atoms per chain to a minimum while also obtaining a non-trivial series of blocks. This simulation contained $N_C=10$ chains.

Buckley, et al. \cite{Buckley2010-ni} reported WAXS measurements of an assortment of PU chemistries involving poly(ethylene adipate) (PEA) soft segments of molar mass 2000 g/mol (approximately $N_S=13$). They used either MDI or 4,4'-dibenzyl di-isocyanate (DBDI) hard segments with $N_H=4$. They use either ethylene glycol (EG), diethylene glycol (DEG), or BDO chain extenders.

\begin{figure}
\includegraphics{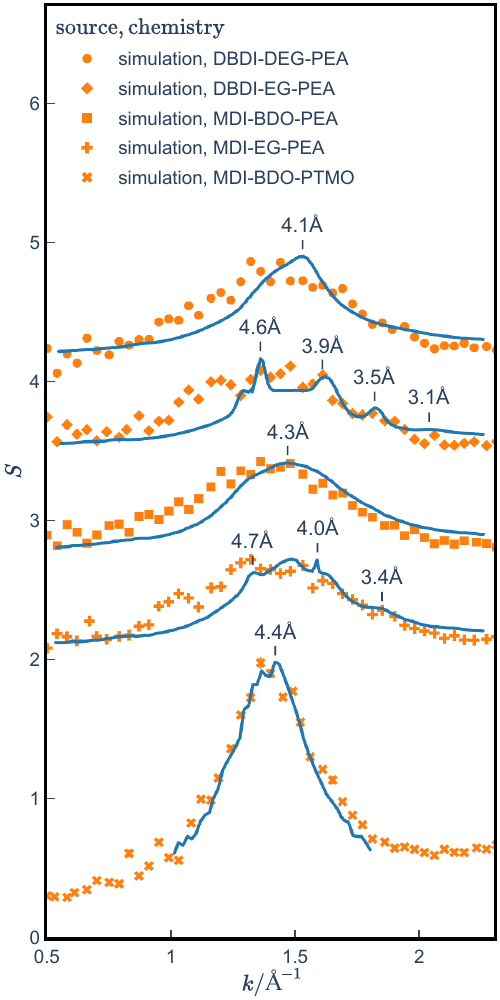}
    \caption{WAXS profiles and corresponding simulation structure factor calculations. Orange symbols are simulation data, blue lines show the corresponding experimental data. Circles, diamonds, squares, and crosses are systems with PEA soft segments reproduced from Buckley, et al. \cite{Buckley2010-ni}, and exes are WAXS data from Sun, et al. \cite{SUN2006650} at 308~K, with permission. Error bars are drawn on simulation data as horizontal lines, but most are smaler than the symbols. Uncertainty estimates were not available for the experimental data, but are likely much smaller.}
    \label{fig:WAXS_chemistries}
\end{figure}

Fig.~\ref{fig:WAXS_chemistries} shows static structure factor calculations for model systems generated according to the above detailed specifications. These calculations are laid over the experimental SAXS profiles of these materials. Note that the scattering profiles of Buckley, et al. were reported as a function of scattering angle $\theta$ and have here been mapped to the wave vector $k$ through $k=(4\pi/\lambda)\sin\theta$, where $\lambda=.82$~\AA{} was the reported x-ray wavelength.

Static structure factors are calculated using the Freud package\cite{freud2020}. The stucture factor calculation takes the form

\begin{equation}
    S(\vec{k})=\frac{1}{N}\sum_{i=0}^N\sum_{j=0}^Ne^{i\vec{k}\cdot \vec{r}_{ij}}.
\end{equation}

Here the vectors $\vec{r}_{ij}$ represent relative positions of each pair of atoms. Vectors $\vec{k}$ are sampled isotropically from a grid defined by the simulation box's reciprocal lattice vectors. Partial structure factors between component $\alpha$ and $\beta$ are computed by limiting $i$ to atoms in $\alpha$ repeat units and $j$ to those in $\beta$ units. Partial structure factors are normalized such that

\begin{equation}
    S(\vec{k})-1=\sum_\alpha\sum_\beta\frac{N_\alpha N_\beta}{N_{total}^2}(S_{\alpha\beta}(k)-1).
\end{equation}

All simulation data points plotted in this work represent an average over time, binned values of k, and two different pseudo-random seeds for initial configuration generation. Simulation data points are marked with vertical lines representing error bars calculated as the standard deviation of the mean of the underlying data. Most error bars are smaller than the plot markers.

\begin{figure}
\includegraphics{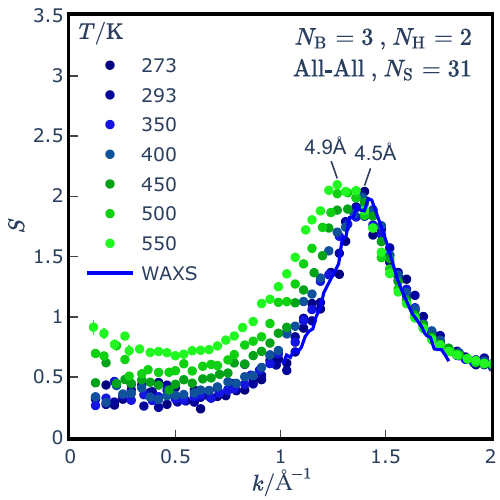}
    \caption{Circles: temperature dependence of the static structure factor for a PTMO-MDI-BDO block copolymer calculated using atomistic simulations. Line: experimental data from a WAXS experiment at 308~K in Sun, et al. \cite{SUN2006650}}
    \label{fig:WAXS_T_sweep}
\end{figure}

Fig.~\ref{fig:WAXS_T_sweep} shows static structure factor calculations for the above specified PTMO model system at a range of temperatures, as well as WAXS measurements reproduced from Sun, et al. The WAXS data has been re-scaled vertically to align with the low-temperature SSF calculations because experimental scattering intensity is reported in counts per second, which is practically an arbitrary unit depending on the measurement protocol. With this adjustment, it is clear that the $k$ value of the first peak and the shape of that peak are reproduced by our model. This is in spite of our small system with only 3 soft blocks per copolymer and only 10 chains per instance. It should be noted that the WAXS measurement reproduced in Fig.~\ref{fig:WAXS_T_sweep} was taken at 308~K, but our SSF calculations don't vary much in the range $1<k<2$~\AA{} and $273<T<350$~K.

Experimental systems can exhibit crystallization\cite{Waletzko2009-ka,Park2021-vq} that is challenging to reproduce in simulation do to large time scales or complicated thermal and strain history during casting. These details can also be lost during heating. Buckley, et al. did not report the temperature at which their measurements were taken. Our PEA simulations for Fig.~\ref{fig:WAXS_chemistries} were carried out at $20^{\circ}$C. The notable horizontal offset between the scattering data and the structure factor for the PEA materials could in part be due to differences in temperature or deformation history.

Nevertheless, Fig.~\ref{fig:WAXS_chemistries} demonstrates that the model generation procedure described above can receive arbitrary chemical components and block length specifications and produce a model that meaningfully represents the material. We can now use these models to systematically investigate the effects of various parameters on the molecular structures.

Several trends are notable in the temperature dependence of $S(k)$ in Fig.~\ref{fig:WAXS_T_sweep}. The peak around $k\sim 1.3$~\AA{} shifts toward lower $k$, corresponding to an increase in specific volume with temperature\cite{Shireen2023-rj,Pugar2020-ba,witzleben2015investigation}. The peak height also increases slightly with increasing temperature. The low-$k$ limit of the SSF also increases significantly with temperature, and undergoes a qualitative change to exhibit a non-monotonic behaviour with a clear upward trend as $k$ approaches 0 at $T\geq500$~K. Sun, et al. also collected SAXS and SANS measurements down to $k=0.01$~\AA, which exhibited a similar increase with decreasing $k$ in the range $0.05<k<0.15$~\AA, and a plateau for $k<0.05$~\AA. Observing this range of $k$ using our model would require a simulation volume with 10 times larger edges, or 1000 times the volume. We are considering a non-isotropic box or coarse graining as strategies to compare our model with such measurements.

\begin{figure}
\includegraphics{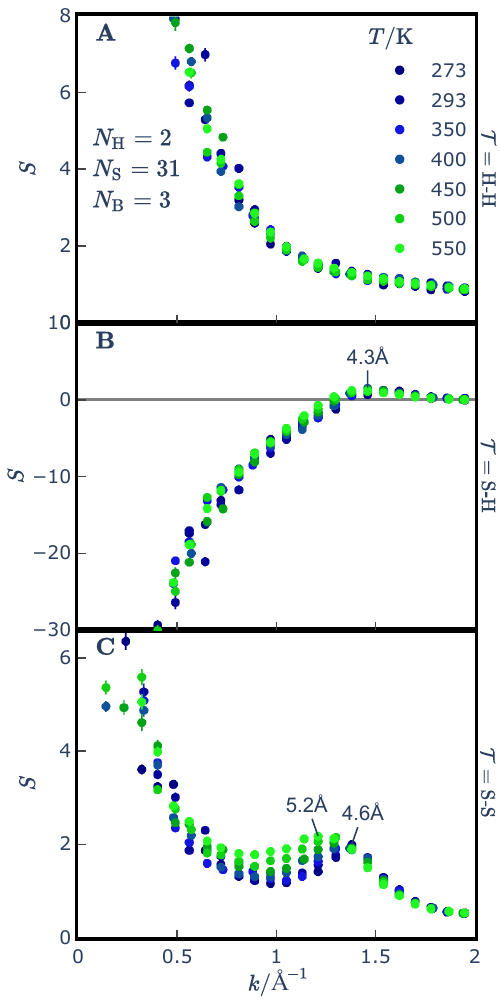}
    \caption{Breakdown of the effect of temperature on partial structure factors calculated using atomistic simulations. A) hard-hard, B) soft-hard C) soft-soft.}
    \label{fig:Partial_T_sweep}
\end{figure}

In simulation, we can decompose the SSF into contributions from each type of interaction between different chemical species in the system. Fig.~\ref{fig:Partial_T_sweep} shows such a breakdown for the same systems as presented inf Fig.~\ref{fig:WAXS_T_sweep}. Fig.~\ref{fig:Partial_T_sweep}A contains the partial structure factors for the interactions between pairs of atoms in hard blocks. Fig.~\ref{fig:Partial_T_sweep}B contains data for all of the interactions between one atom in a soft block and one atom in a hard block. Finally, Fig.~\ref{fig:Partial_T_sweep}B includes pairs of atoms that are both in soft blocks. One notable result in these curves is that the hard-hard and hard-soft structure factors don't seem to change with temperature in the observed range. Secondly, the peak near $k\sim 1.3$ is not visible in the hard-hard curves, only in the hard-soft and soft-soft curves. Finally, the peak in the soft-soft data exhibits the shift toward lower $k$ with increasing $T$. This suggests that the shift toward lower $k$ (indicating a larger structural length scale), is possibly due to increasing volume per soft monomer at increased temperature.

\begin{figure}
\includegraphics{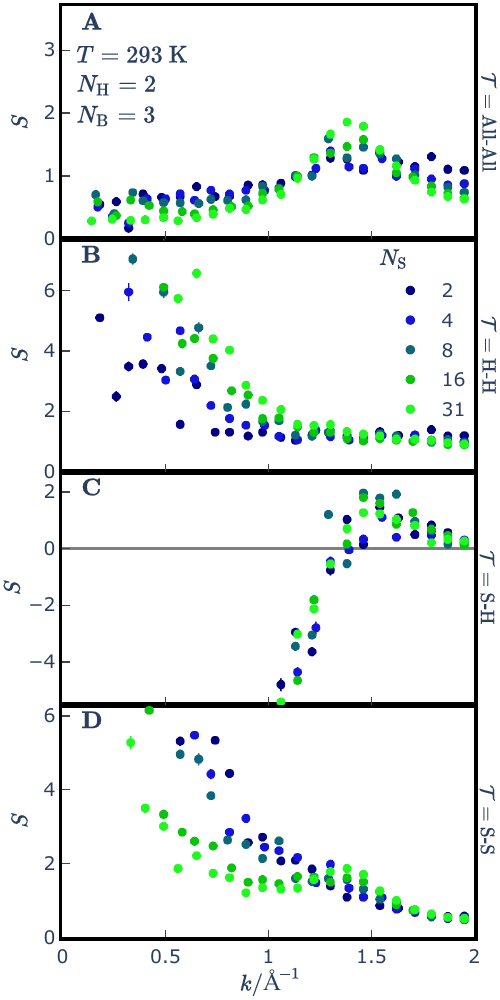}
    \caption{Effect of soft block length on partial structure factors calculated in atomistic simulations. A) All-All, B) hard-hard, C) soft-hard D) soft-soft.}
    \label{fig:Ns_sweep_atoms}
\end{figure}

In Fig.~\ref{fig:Ns_sweep_atoms}, we present the total and partial structure factors with the temperature fixed, but now varying the composition through the number of monomers per soft block $N_S$\cite{Velankar1998-bp,Buckley2017-ya}. The notable trend in the total SSF in Fig.~\ref{fig:Ns_sweep_atoms}A is that the prominent peak when $N_S=31$ seems to decay steadily with $N_S$, and the surrounding observations increase, all the way down to $k$ approaching 0.  The partial structure factors exhibit a variety of trends. The the low-$k$ slope of $S_{HH}$ decreases significantly as $N_S$ decreases. 

\begin{figure}
\includegraphics{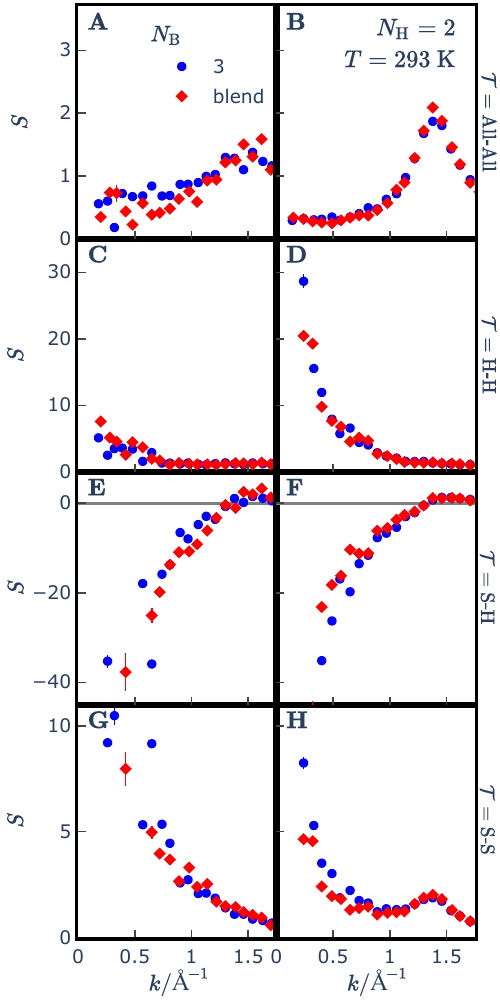}
    \caption{Comparison of partial structure factors for blend vs block copolymer architecture in atomistic simulations with long soft blocks ($N_S=31$ soft repeat units). $N_B=3$ (red diamonds) indicates three soft blocks per chain. $N_B=0$ (blue circles) indicates a 1:1 molar blend of chains composed solely of either soft or hard+extender repeat units.}
    \label{fig:Nb_comparison}
\end{figure}

Figs.~\ref{fig:Nb_comparison} (A,C,E, and G) illustrate the effect of block connectivity. $N_\mathrm{B}$=0 indicates a blend of hard and soft blocks as opposed to a BCP. There is little effect of the block connectivity on the small scale structure around $k\approx1.4$. At low $k$, the copolymer and blend structure factors diverge, suggesting that the architecture has a more significant impact at larger scales, which is not surprising as separate hard and soft chains will prefer to separate. It is notable that this low-$k$ separation is not observed in the total structure factor, but only the hard-hard and soft-soft partials.

Fig.~\ref{fig:Nb_comparison} (B,D,F, and H) also explore the effect of block connectivity, but this time in the context of short soft blocks. In this case the distinction between the blend and the BCP is erased. This result is consistent with the concept that short chains are less prone to phase separate, so the large length scale features don't appear, regardless of the connectivity.

\section{Conclusion}

In this work we have implemented and validated a platform for automatic generation of atomistic models of arbitrary polyurethane block copolymers and blends. We build on the RadonPy package\cite{hayashi2022radonpy} by introducing a strategy to represent urethane between components. We articulated the general architectural pattern of a PU BCP in a format that enabled heirarchical construction of the hard and soft blocks, then a full chain. We developed a routine for efficiently computing bespoke charge distributions automatically for each type of monomer depending on it's neighboring monomers. We compared static structure factor calculations for generated models to experimental WAXS data\cite{SUN2006650,Buckley2010-ni} for corresponding materials. We then demonstrated the utility of such a model generator to finely control individual design parameters by surveying the dependence of structure factor on temperature, soft block length, and block connectivity. We noted that some trends were only apparent in the partial structure factors, which are uniquely provided by simulations.

The workflow demonstrated here is not without its challenges. With any atomistic simulation, the accessible time and length scales are limited. In the context of scattering data, it would be useful to be able to model length scales covered by small angle\cite{SUN2006650,Dhollander2010-dc,witzleben2015investigation,Wang2022-qm,Miller1984-pl,Leung1985-ok} as well as wide angle x-ray scattering. These measurements provide information about phase separated microstructures, which are characterized by length scales 10-100 times larger than those contained in the simulations presented here. The corresponding volume increase would cost thousands of times the computational resources as this work. Time scales are also a limitation, because these materials can exhibit partial crystallization\cite{Waletzko2009-ka} and glassy dynamics\cite{Xia2018-fy,Pugar2020-ba}, which would also require orders of magnitude longer simulations to confidently study. The class of models constructed here also do not fully account for reaction kinetics\cite{Speckhard1987-kj,lee2011plastic,in2003temperature}, either during synthesis or under measurement conditions. Synthesis details could have an effect on the distribution of block lengths and the presence of unreacted components, which could impact morphology. Some bonds are affected by high temperature, either shifting the appropriate force field parameters, or allowing bonds to break completely, which limits the valid temperature range of a procedurally generated model in a manner that is challenging to predict.

There are several directions for further development of the work presented here. The block construction routine could be straightforwardly modified to represent other copolymer templates. A more complex feature to implement would be a method to encode a polymerization reaction and find groups in components that can undergo such a reaction. Such a feature would vastly widen the applicability of this work to systematic modelling of polymers, but would require non-trivial pattern matching logic. A further enhancement would be to enable construction of branched polymers. The polymerization functions in RadonPy presently only support two link sites per monomer, but the data structures for labeling such sites would be easy to modify to allow an arbitrary number of crosslinks from a repeat unit. A subtle shortcoming of RadonPy is that during polymerization, each time a repeat unit is added, the entire polymer data structure is copied, require to $O(N^2)$ time for initialization, which makes longer chains intractable to simulate. This limitation could be overcome by a suitable modification to RadonPy's function for combining molecules. While the atomic charges are determined by quantum mechanical simulations, the bonded and Van der Waals interaction parameters still set using typical atom types and tabulated force fields. Methods exist for extracting at least some interaction parameters from quantum mechanical simulations as well\cite{horton2019qubekit}.

Aside from technical improvements, several further studies are now possible with the methods demonstrated here. The results of these simulations are readily used as training data for machine learning models\cite{Pugar2020-ba,Shireen2022-hm,Pugar2022-cg,Ethier2023-qd}. Ready access to detailed trajectories and additional samples could augment efforts to interpret and explain such models. The robustness of this workflow allows it to be run automatically within a larger scheme for active learning or property optimization\cite{Weeratunge2022-dm}. The data generated from these atomistic simulations could be used as bottom-up ground truth data for development of coarse-grained models\cite{Weeratunge2023,Xia2018-fy,Dunbar2020-nm,Giuntoli2021-hu,Park2021-vq,Shireen2022-hm} which could mitigate the constraints on time and length scales. These are just a few examples of how an interoperable workflow for procedurally generating atomistic models could couple to multi-scale studies of structure-property relationships. The models generated here are also readily analyzed to measure rheological\cite{Velankar2000-dd,Velankar1998-bp}, thermal\cite{in2006temperature}, and electrical\cite{hayashi2022radonpy} properties. Another important aspect of modern polymer design is the effect of confinement, which is vital for studying porous media or nanofluidics\cite{sg1, sg2, sg3}. The polymer models generated using the method discussed here could be included in more complex simulations with details such as walls or flow to study these applications.

\section{Funding}

This research is supported by the Commonwealth of Australia as represented by the Defence Science and Technology Group of the Department of Defence through the multi-disciplinary materials sciences stream of the Next Generation Technologies Fund.
 
\bibliography{main}

\end{document}